\documentclass[prb, twocolumn, superscriptaddress, showpacs]{revtex4}  
\usepackage{amssymb}
\usepackage{amsmath}
\usepackage{bbm}
\usepackage{bm}
\usepackage{graphicx}
\newcommand{\kk}{{\bm k}} 
\newcommand{\tp}{t^{\zz}} 
\newcommand{\pp}{ \text{\raisebox{1pt}{${}_{|\!|}$}}   }
\newcommand{\tpp}{t^{\pp}} 
\newcommand{\zz}{ \text{\raisebox{1pt}{${}_{\perp}$}}   }

\begin{document}

\author{Martin Eckstein}
\affiliation{Max Planck Research Department for Structural Dynamics, University of Hamburg-CFEL, Hamburg, Germany}
\author{Philipp Werner}
\affiliation{Department of Physics, University of Fribourg, 1700 Fribourg, Switzerland}

\title{Nonequilibrium dynamical mean field simulation of inhomogeneous systems}

\date{\today}

\begin{abstract}
We extend the nonequilibrium dynamical mean field (DMFT) formalism to inhomogeneous systems by adapting the ``real-space" DMFT method to Keldysh Green's functions. Solving the coupled impurity problems using strong-coupling perturbation theory, we apply the formalism to homogeneous and inhomogeneous layered systems with strong local interactions and up to 39 layers.   We study the diffusion of doublons and holes created by photo-excitation in a Mott insulating system, the 
time-dependent build-up of the 
polarization and 
the
current induced by a linear voltage bias across a multi-layer structure, and the photo-induced current in a Mott insulator under bias. 
\end{abstract}

\pacs{71.10.Fd}

\maketitle

\hyphenation{}

%\section{Model}

\section{Introduction}

The study of nonequilibrium phenomena in correlated lattice systems has become an active research field due to experimental progress on several fronts. In cold atom systems, the interaction and bandwidth can be controlled via Feshbach resonances and the depth of the lattice potential, respectively, while the effect of external fields can be mimicked by shaking or tilting the optical 
lattice.\cite{Greiner2002, Strohmaier2008, Liniger2007,Struck2011,Simon2011} 
This allows to investigate quench-dynamics or field-driven effects in systems which may be viewed as ideal realizations of the simple model Hamiltonians typically considered in theoretical studies. On the other hand, advances in ultra-fast laser science have made it possible to perturb a correlated material with a strong pulse and track the time-evolution of the system with the (femto-second) time resolution needed to 
observe 
intrinsically electronic processes.\cite{Cavalieri2007, Wall2011} Such experiments can provide new insights into the nature of correlated states of matter and may even lead to the discovery of `hidden phases', i.e. long-lived transient states that cannot be accessed via a thermal pathway. 

Stimulated by these developments, a growing theoretical effort is aimed at describing and understanding the nonequilibrium properties of correlated lattice systems. Given the complexity of the task, much of this work has focused on the simplest relevant model, the one-band Hubbard model, which describes electrons 
that can hop
between nearest-neighbor sites of some lattice with hopping amplitude $t$,
and 
interact on-site with a repulsion energy $U$. A method which is well-suited to capture strong local correlation effects is dynamical 
mean-field 
theory (DMFT),\cite{Metzner1989, Georges1996} 
and this formalism can be
extended to nonequilibrium systems in a rather straightforward manner.\cite{Schmidt2002, Freericks2006} 
Over the last few years, nonequilibrium DMFT has been used in a large number of theoretical studies of the nonequilibrium dynamics in homogeneous bulk systems, including
interaction-quenches,\cite{Eckstein2009, Eckstein2010}
dc-field driven Bloch oscillations\cite{Freericks2006,Eckstein2011bloch} or
insulator-to-metal transitions\cite{Eckstein2010breakdown, Eckstein2013breakdown}
(and the related phenomenon of dimensional reduction\cite{Aron2012}),
photo-doping,\cite{Eckstein2011} ac-field induced band-flipping,\cite{Tsuji2011} 
and nonequilibrium phase transitions from antiferromagnetic to paramagnetic states.\cite{Werner2012, Tsuji2012} 
While connecting these results to actual experiments is difficult because of the idealized set-up in the model calculations, 
they have provided important insights into the relaxation dynamics of purely electronic systems, and the associated time-scales 
and trapping phenomena. 

One step towards more realistic model calculations is to switch from infinitely extended, homogenous systems to a description 
which allows for a spatial variation in the model parameters. 
In equilibrium, the inhomogeneous DMFT approach\cite{Potthoff1999,Freericks2004} allows, for example, to describe some effect
of the trapping potential in cold-atom experiments,\cite{Helmes2008a,Gorelik2010} or correlation effects in artificially designed 
heterostructures.\cite{Potthoff1999,Freericks2004,Helmes2008b}  
In a direct generalization of this real-space DMFT to nonequilibrium, 
one would have to store and manipulate Green's functions $G_{ij}(t,t')$ 
which depend on two space arguments $i,j$ and two time arguments. Decoupling of space and time is no longer possible, 
neither by introducing momentum-dependent Green's functions $G_\kk(t,t')$ (as in homogeneous nonequilibrium 
DMFT), nor by using frequency-dependent  Green's functions  $G_{ij}(\omega)$ (as in inhomogeneous equilibrium 
DMFT). The fully inhomogeneous set-up would thus require a prohibitively large amount of memory for most  
applications. However, the problem turns out to be numerically tractable for a simpler layered geometry, which is still relevant 
for many applications. Here one considers a system
in which the properties can change as a function of the lattice position in one direction, while being homogeneous in the 
$d-1$ other dimensions. 
For example, such an extension allows to deal with
surface phenomena in condensed matter systems, such as the propagation 
of excitations from the surface of a sample into the bulk
(which has been looked at recently within a time-dependent Gutzwiller approach\cite{Andre2012}). 
In this context it is important to mention that pump-probe experiments often 
excite only a thin surface layer, such that interesting phenomena must be inferred
by subtracting the bulk-signal, based on some assumptions about the penetration depth of the pump 
pulse. The layer description also naturally lends itself to the study of interfaces and heterostructures.\cite{Ohtomo2004, Okamoto2004} 
The latter are at present the subject of extensive research, and 
experimental results on ultra-fast photo-induced metal-insulator transitions in heterostructures have recently been 
published.\cite{Caviglia2012}

In this paper, we discuss and test an implementation of  the
nonequilibrium DMFT formalism for inhomogeneous, layered structures. 
This formalism is an adaptation of the equilibrium ``real-space" DMFT method developed by Potthoff, Nolting, Freericks and 
others.\cite{Potthoff1999, Freericks2004} We discuss the formalism and the techniques used for solving the DMFT equations 
in Sec.~\ref{Sec_method}, and illustrate the versatility of the approach in Sec.~\ref{Sec_results} with several test calculations involving electric field pulse excitations of correlated layers or heterostructures. Section~\ref{Sec_conclusions} gives a brief conclusion and outlook.

\section{Model and Method}
\label{Sec_method}

The approximate DMFT treatment of layered structures assumes a local self-energy for each layer and maps the system to 
an effectively one-dimensional model subject to a self-consistency condition. An efficient strategy for solving the DMFT equations, 
which involves a partial Fourier transformation of the Green's functions
with respect to the transverse space directions,  
has been proposed by Potthoff and Nolting,\cite{Potthoff1999} and a detailed description of the equilibrium implementation of this 
so-called ``real-space DMFT" can be found 
in work by Freericks.\cite{Freericks2004, Freericks_book}
Here, we adopt this technique to nonequilibrium systems in order to describe pulse-excitations of surfaces or heterostructures, 
as well as transport through correlated thin films, using the nonequilibrium DMFT formalism.\cite{Schmidt2002, Freericks2006} 

\begin{figure}[t]
\begin{center}
\includegraphics[angle=0, width=0.9\columnwidth]{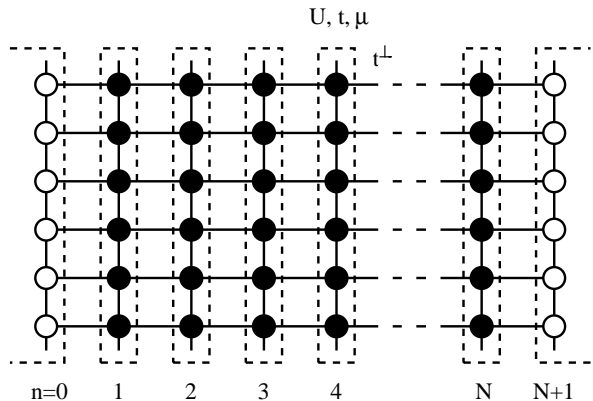}
\caption{Illustration of the layer set-up with $N$ correlated layers (full dots), intra-layer hopping $t$,  inter-layer hopping 
$t^\perp$, interaction $U$, and chemical potential $\mu$ (all these parameters can be layer- and 
time-dependent). The boundary condition is given either by coupling to some noninteracting equilibrium bath, by a vacuum 
(no hopping into the boundary layers), or by repeating the hybridization functions of the surface layers. 
}
\label{fig_layer}
\end{center}
\end{figure} 

We consider a Hubbard model with $N$ layers, connected by an interlayer hopping $\tp$, and either ``vacuum", ``lead" or ``bulk" boundary 
conditions applied to the left ($n=1$) and right ($n=N$) surface layers (see Fig.~\ref{fig_layer}). Here,  ``vacuum" means no hopping to the 
boundary layer, ``lead" means we impose some equilibrium DMFT solution in the boundary layer, and ``bulk" means that the solution on the 
surface layer is repeated periodically. 
The corresponding Hamiltonian is given by 
\begin{align}
H =
&\sum_{n=1}^{N} 
\Big[
-\sum_{ij\sigma}
\tpp_{n,ij}
\,c_{i,n,\sigma}^\dagger c_{j,n,\sigma}
+
\epsilon_{\text{loc},n}
\sum_{i\sigma}
c_{i,n,\sigma}^\dagger c_{i,n,\sigma}
\Big]
\nonumber
\\
&+
\sum_{n=1}^{N} 
\sum_{i}
U_n\,
c_{i,n,\uparrow}^\dagger c_{i,n,\uparrow}
c_{i,n,\downarrow}^\dagger c_{i,n,\downarrow}
\nonumber
\\
&+
\sum_{n=1}^{N-1}
\sum_{i\sigma}
\Big(
-\tp_{n}
c_{i,n,\sigma}^\dagger
c_{i,n+1,\sigma}+ h.c.
\Big)
+ b.t.,
\label{hamiltonian}
\end{align}
where $c_{i,n,\sigma}^\dagger$ creates an electron on lattice site $i$ in layer $n$, 
$U_n$ is the on-site Coulomb interaction in layer $n$, $\epsilon_{\text{loc},n}$ is a layer-dependent  on-site potential, 
and $\tpp$ and $\tp$ denote the hopping within the layers and between the 
layers, respectively.
%%NOTE: in the code, t_n is the "right-hopping" tR[n]= t_{n,n+1} which is provided by the input file
The term ``$b.t.$'' summarizes the boundary terms as described above.
All parameters can depend both on time and on the layer index, which will mostly not be shown explicitly in the following. 
In the actual implementation, each layer corresponds to a $d$-dimensional hypercubic lattice with lattice spacing $a$, and we present 
results for  $d=1$. We will later switch to the Fourier transformation with respect to the intra-layer coordinate, $c_{j,n,\sigma}= 
\frac{1}{\sqrt{N_\kk}}\sum_\kk e^{i\kk {\bm r}_j/a }c_{\kk,n,\sigma}$. The intra-layer hopping Hamiltonian becomes 
$\sum_{\kk\sigma} \!\epsilon_{n,\kk} \,c_{\kk,n,\sigma}^\dagger c_{\kk,n,\sigma}$, with the dispersion 
$\epsilon_{n,\kk}=-\sum_i t^\pp_{n,ij} e^{i\kk ({\bm r}_j-{\bm r}_i)/a }$.

External electromagnetic fields are included in Eq.~(\ref{hamiltonian}) via the Peierls substitution: We consider electric fields 
${\bm E} \equiv({\bm E}^\pp_n,E^\zz_n)$ that depend only on the layer coordinate, and let ${\bm E}^\pp_n$ and $E^\zz_n$ 
denote the parallel field component in layer $n$ and the perpendicular field-component between layer $n$ and $n+1$, 
respectively. Units for the fields are taken as  
$[t]/ea$ for ${\bm E}^\pp$ and $[t]/ea_{\zz}$ for $E^\zz$, 
where $[t]$ is a unit of 
energy, $a_\zz$ is the spacing between layers, and $-e$ is the electron charge. In  a gauge where also the scalar potential 
$\phi_n$ and vector potential ${\bm A} \equiv({\bm A}^\pp_n,A^\zz_n)$ depend on the layer only, we then have
${\bm E}^\pp_n = - \partial_t {\bm A}^\pp_n$
and 
${\bm E}^\zz_n = - \partial_t {\bm A}^\zz_n-(\phi_{n+1}-\phi_n)$,
and the Peierls substitution gives
\begin{align}
&\epsilon_{n,\kk}
=
\tilde 
\epsilon_{
n,\kk + {\bm A}^\pp_n},
\\
&t^{\zz}_n 
=
\tilde t^{\zz}_n 
\exp\big( i A^{\zz}_{n} \big),
\\
&\epsilon_{\text{loc},n} 
=
\tilde\epsilon_{\text{loc},n} - \phi_n,
\end{align}
where quantities with tilde correspond to zero field. 
Also for fields perpendicular to the layer it is often convenient to use a gauge with zero scalar potential.

Nonequilibrium DMFT provides a set of equations for the space- and time-dependent Green's functions $G_{i,n;j,m}(t,t') 
= - i \langle \mathcal{T}_\mathcal{C} c_{i,n,\sigma}(t) c_{j,m,\sigma}^\dagger(t')\rangle$. Here $t$ and $t'$ lie on the 
L-shaped Keldysh contour $\mathcal{C}$, and $\mathcal{T}_\mathcal{C}$ is the contour-ordering operator. The notation 
for contour-ordered Green's functions and their inverse operators is adopted from Ref.~\onlinecite{Eckstein2010}.
The functions $G_{i,n;j,m}(t,t')$ are obtained from the lattice Dyson equation with a local but layer-dependent self-energy 
$\Sigma_n(t,t')$, which is computed from an effective impurity model (see below; for simplicity we omit a possible 
dependence of local quantities on spin). Due to the translational invariance within the layers, one 
can perform a Fourier transformation in the transverse directions and introduce the momentum-dependent Green's functions 
$G_{\kk;n,m}(t,t') = - i \langle \mathcal{T}_\mathcal{C} c_{\kk,n,\sigma}(t) c_{\kk,m,\sigma}^\dagger(t')\rangle$.
The Dyson equation then decouples for each $\kk$, and one has the following matrix expression
for the $N\times N$ matrices $(G_{\kk})_{n,m}\equiv G_{\kk;n,m}$,
\begin{multline}
\label{dyson-chain}
(G^{-1}_{\kk})_{m,n}=(i\partial_t+\mu-\epsilon_{\text{loc},m}-\epsilon_{\kk,m}-\Sigma_m)\delta_{m,n}
\\
-\tp_m\delta_{m+1,n}-(t^{\zz}_{m-1})^*\delta_{m-1,n},
\end{multline}
which is equivalent to the Dyson equation for a one-dimensional chain with sites $m=1,...,N$.
The local Green's function on layer $n$ is then computed from $G_n=\frac{1}{N_\kk}\sum_\kk (G_\kk)_{n,n}$,
and hence we only need the diagonal elements $(G_{\kk})_{n,n}$ of the momentum-dependent Green's function. 
These can be evaluated using the following formulas for the inverse of a tri-diagonal matrix, 
\begin{align}
&M^{-1}=\left(
\begin{array}{cccccc}
z-a_1 & b_1 & & & &\\
b_1^* & z-a_2 & b_2&&&\\
 & & \ldots &&&\\
 & & &\ldots&b_{n-1}&\\
 & &&b_{n-1}^* & z-a_n&\\ 
\end{array}
\right),\\
&M_{11}=\frac{1}{z-a_1-\frac{|b_1|^2}{z-a_2-\frac{|b_2|^2}{z-a_3 - \ldots}}},\\
&M_{22}=\frac{1}{z-a_2-\frac{|b_1|^2}{z-a_1}-\frac{|b_2|^2}{z-a_3-\frac{|b_3|^2}{z-a_4 - \ldots}}},\\
&M_{33}=\frac{1}{z-a_3-\frac{|b_2|^2}{z-a_2-\frac{|b_1|^2}{z-a_1}}-\frac{|b_3|^2}{z-a_4-\frac{|b_4|^2}{z-a_5 - \ldots}}},\\
&\ldots\nonumber .
\end{align}
Explicitly, one finds
\begin{align}
(G_\kk)_{n,n}
&=\frac{1}{g^{-1}_{\kk,n}
-\Delta^L_{\kk,n-1}
-\Delta^R_{\kk,n+1}
},
\label{Gknn}
\\
g^{-1}_{\kk,n}
&=i\partial_t+\mu-\epsilon_{\text{loc},n}-\epsilon_{\kk,n}-\Sigma_n,
\label{gkn1}
\end{align}
where $g_{\kk,n}$ is the Green's function corresponding to an isolated layer, 
and we have introduced  the products
\begin{align}
\Delta^L_{\kk,n-1}(t,t')
&=t^{\zz*}_{n-1}(t)\, G_{\kk,n-1}^{[n]}(t,t')\, t^\zz_{n-1}(t')
\label{Delta_L_product}
\\
&\equiv
t^{\zz*}_{n-1}\ast G_{\kk,n-1}^{[n]}\ast t^\zz_{n-1}
\\
\Delta^R_{\kk,n+1}(t,t')
&=
t^\zz_n(t) \,G_{\kk,n+1}^{[n]}(t,t') \, t^{\zz*}_n(t'),
\label{Delta_R_product}
\end{align}
which involve the Green's functions $G_\kk^{[n]}$ for
the ``chain'' (Eq.~(\ref{dyson-chain})) with site $n$ removed.  
The Green's functions $G_\kk^{[n]}$ satisfy equations analogous to Eq.~(\ref{Gknn}),
such that we obtain for the hybridizations $\Delta^L_{\kk,n}$ and $\Delta^R_{\kk,n}$ 
\begin{align}
\Delta^L_{\kk,n} 
&= 
 t^{\zz*}_{n} \ast 
 \frac{1}{g_{\kk,n}^{-1}-\Delta^L_{\kk,n-1}}  
 \ast t^\zz_{n} \label{Delta_L},
 \\ 
\Delta^R_{\kk,n}
&=
t^\zz_{n-1} \ast 
\frac{1}{g_{\kk,n}^{-1}-\Delta^R_{\kk,n+1}} 
\ast t^{\zz*}_{n-1},
\label{Delta_R}
\end{align}
for layers $n=1, \ldots, N$. The boundary conditions read $\Delta_{\kk,n}=0$ (``vacuum") or $\Delta_{\kk,n}=\Delta_{\kk,\text{lead}}$ 
(``lead") for $n=0, N+1$. The ``bulk" boundary condition is 
$\Delta^R_{N+1}=\Delta^R_{N}$,  $\Delta^L_{0}=\Delta^L_{1}$.
Once the $\Delta^L_{\kk,n-1}$ and $\Delta^R_{\kk,n+1}$ for a given layer $n$ have been updated, one computes $(G_\kk)_{n,n}$ 
using Eq.~(\ref{Gknn}), and determines the hybridization function $\Lambda_n=\Lambda_n[G_n]$ of the impurity model by solving 
the impurity Dyson equation,
\begin{equation}
G_{n}=\frac{1}{N_\kk}\sum_{k} (G_\kk)_{n,n}\equiv \frac{1}{i\partial_t+\mu-\epsilon_{\text{loc},n}-\Sigma_n-\Lambda_n}.\label{def_lambda}
\end{equation}
The solution of the impurity problem (in the present case, we use the non-crossing approximation (NCA)\cite{Keiter1973, Eckstein2010nca} as impurity solver) yields an updated $G_n$ and $\Sigma_n$. 

A self-consistent solution on all layers can hence be obtained by the ``zipper algorithm":\cite{Freericks2004}
\begin{align}
&
\left.
\begin{array}{llll}
n=1 \hspace{0.5cm}&\downarrow\hspace{2cm}   & \uparrow \hspace{2cm}  & \downarrow\\
n=2 &\downarrow   & \uparrow &\downarrow \\
\ldots &\downarrow \Delta^L_n  & \uparrow \Delta^R_n, G_n, \Lambda_n & \downarrow \Delta^L_n, G_n, \Lambda_n\\
&\downarrow   & \uparrow & \downarrow \\
n=N & \downarrow  & \uparrow & \ldots \\
\end{array}
\right.\nonumber
\end{align}
where we start for example with $\Lambda_n=\Lambda_\text{bulk}$, $\Sigma_n=\Sigma_\text{bulk}$, for $n=1,\ldots, N$, $\Delta^L_1=\Delta^R_N=0$, 
and then update $\Delta^L_n$ using Eq.~(\ref{Delta_L}) from $n=1$ to $N$. On the way back, we use Eq.~(\ref{Delta_R}) to update 
$\Delta^R_n$, from $n=N$ to $n=1$, and at the same time compute $G_n$, $\Lambda_n$ and $\Sigma_n$ for each of these $n$, and so on.

Equations (\ref{Gknn})-(\ref{def_lambda}) are integral-differential equations on the Keldysh contour. 
Following the strategy outlined in Ref.~\onlinecite{Eckstein2011}, we can cast these equations in a form 
that can conveniently be handled  by numerically stable ``time-stepping'' procedures for the propagation 
of Green's functions in real time. 
Defining the variables $\xi_{\kk,n}=\epsilon_{\kk,n}+\Delta^L_{\kk,n-1}+\Delta^R_{\kk,n+1}$ and 
$Z_n=[i\partial_t+\mu-\epsilon_{\text{loc},n}-\Sigma_n]^{-1}$, one can write Eqs.~(\ref{Gknn}) and (\ref{def_lambda}) 
in the form (dropping for simplicity the index $n$ everywhere)
\begin{align}
&[Z^{-1}-\xi_\kk]*G_\kk=I, & I=G_\kk*[Z^{-1}-\xi_\kk],\label{eqkk}\\
&[Z^{-1}-\Lambda]*G=I, & I=G*[Z^{-1}-\Lambda].\label{eq_lambda}
\end{align}
By summing Eq.~(\ref{eqkk}) over $\kk$ and comparing with Eq.~(\ref{eq_lambda}), one finds 
$G^{(1)}\equiv\sum_\kk \xi_\kk * G_\kk=\Lambda*G$ and ${G^{(1)}}^\dagger\equiv\sum_\kk G_\kk*\xi_\kk=G*\Lambda$. 
We next take the second of Eqs.~(\ref{eq_lambda}) and multiply from the right with $Z$. This leads to 
\begin{equation}
[I+G*\Lambda]*Z=[I+{G^{(1)}}^\dagger]*Z=G,\label{eq_Z}
\end{equation}
which we can solve for $Z$ (after having evaluated ${G^{(1)}}^\dagger=G*\Lambda$). Multiplying the first of Eqs.~(\ref{eqkk}) 
from the left with $Z$ gives
\begin{equation}
[I-Z*\xi_\kk]*G_\kk=Z,\label{eq_Gk}
\end{equation}
which we can solve for $G_\kk$.
From the first Eq.~(\ref{eq_lambda}), we also get $[I+\Lambda*G]*\Lambda=Z^{-1}*G*\Lambda=\sum_\kk Z^{-1}*G_\kk*\xi_\kk$. 
But from the first Eq.~(\ref{eqkk}), $Z^{-1}*G_\kk=I+\xi_\kk *G_\kk$, so 
\begin{align}
&[I+\Lambda*G]*\Lambda=[I+G^{(1)}]*\Lambda=G^{(2)},\label{eq_lambda_final}
\end{align}
where $G^{(2)}=\sum_\kk(\xi_\kk+\xi_\kk*G_\kk*\xi_\kk)$. 
The solution of Eq.~(\ref{eq_lambda_final}) yields $\Lambda$.

To solve Eqs.~(\ref{Delta_L}) and (\ref{Delta_R}), we first compute $g_{\kk,n}$ [Eq.~(\ref{gkn1})]
by solving 
\begin{align}
& [I-Z_n*\epsilon_\kk]*g_{\kk,n}=Z_n.\label{eq_gk}
\end{align}
With the shorthand notation 
$\tilde \Delta^L_{\kk,n}=(G_\kk^{[n+1]})_n$
and 
$\tilde \Delta^R_{\kk,n}=(G_\kk^{[n-1]})_n$,
we rewrite Eqs.~(\ref{Delta_L}) and (\ref{Delta_R}) as 
\begin{align}
&[I-g_{\kk,n}\ast t^{\zz *}_{n-1} \ast \tilde\Delta^L_{\kk,n-1} \ast t^\zz _{n-1}]*\tilde\Delta^L_{\kk,n}=g_{\kk,n},\label{eq_Delta_L}
\\
&[I-g_{\kk,n}\ast t^\zz_n \ast \tilde\Delta^R_{\kk,n+1} \ast t^{\zz *}_n]*\tilde\Delta^R_{\kk,n}=g_{\kk,n}.\label{eq_Delta_R}
\end{align}

Equations (\ref{eq_Z}), (\ref{eq_Gk}), (\ref{eq_lambda_final}), (\ref{eq_gk}), (\ref{eq_Delta_L}) and (\ref{eq_Delta_R}),
are all of the form $[I+A]*X=B$ and have to be solved for $X$. This is an integral equation of the Volterra type, which 
is well behaved and which we solve using the techniques described in Ref.~\onlinecite{Eckstein2010}. The solution 
can be obtained by successively increasing the maximum time in a step by step manner, thereby not modifying an 
already converged solution at earlier times. 

In summary, at a given time-step, we perform the following 
calculations in layer $n$:
\begin{enumerate}
\item For given $\Lambda_n$, solve impurity problem (NCA equations) to obtain $G_n$.
\item Evaluate ${G^{(1)}_n}^\dagger=G_n*\Lambda_n$, solve Eq.~(\ref{eq_Z}) for $Z_n$.
\item For each $\kk$-point,
\begin{itemize}
\item solve Eq.~(\ref{eq_gk}) for $g_{\kk,n}$,
\item solve equations of the type (\ref{eq_Delta_R}) and (\ref{eq_Delta_L}) to get the new $\tilde \Delta_{\kk,n}^R$ or $\tilde \Delta_{\kk,n}^L$,
and compute  $\Delta_{\kk,n}^R$ or $\Delta_{\kk,n}^L$ from Eqs.~(\ref{Delta_L_product}) and (\ref{Delta_R_product}) (depending on the 
direction of the sweep),
\item define $\xi_{\kk,n}=\epsilon_\kk+\Delta^L_{\kk,n-1}+ \Delta^R_{\kk,n+1}$ and solve Eq.~(\ref{eq_Gk}) for $G_{\kk,n}$.
\end{itemize}
\item Having obtained $\xi_{\kk,n}$ and $G_{\kk,n}$ for all $\kk$-points, calculate $G^{(1)}_n$ and $G^{(2)}_n$.
\item Solve Eq.~(\ref{eq_lambda_final}) to obtain the new $\Lambda_n$.
\end{enumerate} 
Then we move to the next layer, where we repeat the same cycle, zipping back and forth until convergence is reached.
Only a few cycles are needed for convergence, since a very good starting point is  obtained by extrapolating the 
Green's functions from earlier times.

Depending on the application, it may be desirable to include a dissipation mechanism which allows to remove energy injected into the system by a quench or external field. In Ref.~\onlinecite{Eckstein2012photodoping} we have briefly described how one can locally couple a phonon bath with given temperature. Let us discuss now how such a bath can be incorporated into the ``zipper algorithm". In our approximation, the electronic self-energy on layer $n$ is the sum of an electronic contribution, $\Sigma_U[G_n]$, and of a bath contribution $\Sigma_\text{diss}[G_n]$. As in the case without bath (Eq.~(\ref{def_lambda})), $\Sigma_U[G_n]$ is obtained from the solution of the impurity problem with hybridization $\Lambda_n$: $G_n=G_n[\Lambda_n]$, with 
\begin{equation}
G_n=\frac{1}{i\partial_t+\mu-\Sigma_U[G_n]-\Lambda_n}.
\label{G_n_Sigma_U}
\end{equation}
The bath contribution is approximated by the lowest order Holstein-type electron-phonon diagram: 
\begin{equation}
\Sigma_\text{diss}[G_n]=\lambda G_n(t,t')D(t,t'),
\end{equation}
with $D(t,t')=-i\text{Tr}[T_\mathcal{C}\exp (-i\int_\mathcal{C}dt \omega_0 b^\dagger b)b(t)b^\dagger(t')]/Z$ the equilibrium boson propagator for boson frequency $\omega_0$ and coupling strength $\lambda$.
Therefore, in Eqs.~(\ref{Gknn}) and (\ref{gkn1}), which relate the momentum dependent lattice Green's function to the self-energy, we have to replace $\Sigma_n$ by $\Sigma_U[G_n]+\Sigma_\text{diss}[G_n]$, or equivalently $\epsilon_\kk$ by $\epsilon_\kk+\Sigma_\text{diss}[G_n]$.\cite{Eckstein2012photodoping}

In practice, we define $Z_\text{latt}=[i\partial_t+\mu-\epsilon_{\text{loc},n}-\Sigma_U-\Sigma_\text{diss}]^{-1}$ (dropping the layer-index $n$), so that Eq.~(\ref{G_n_Sigma_U}) becomes $G=1/(Z^{-1}_\text{latt}-\Lambda_\text{latt})$, with $\Lambda_\text{latt}=\Lambda-\Sigma_\text{diss}$. We may then repeat the derivation of Eqs.~(\ref{eq_lambda})-(\ref{eq_lambda_final}) with the substitution $\Lambda\rightarrow\Lambda_\text{latt}$, $Z\rightarrow Z_\text{latt}$, {\it i.e.}, given $\Lambda_\text{latt}$ and $G$, $Z_\text{latt}$ is computed from $[I+{G^{(1)}}^\dagger]*Z_\text{latt}=G$ (with ${G^{(1)}}^\dagger=G*\Lambda_\text{latt}$), then a new $\Lambda_\text{latt}$ is obtained from the solution of $[I+G^{(1)}]*\Lambda_\text{latt}=G^{(2)}$. Finally, $\Lambda=\Lambda_\text{latt}+\Sigma_\text{diss}$ is used as input for the impurity solver.

\begin{figure}[t]
\begin{center}
\includegraphics[angle=0, width=\columnwidth]{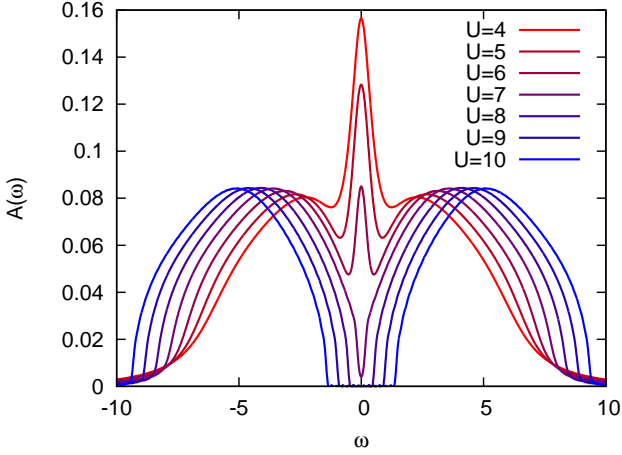}
\caption{Equilibrium spectral functions for an infinite system of 1-$d$ layers, at $\beta=5$ and indicated values of the interaction strength. The DMFT solution has been obtained with an NCA impurity solver, which yields reliable results in the insulating phase. 
}
\label{equilibrium}
\end{center}
\end{figure}

\section{Results}
\label{Sec_results}

\subsection{Test of the implementation}

In this work we will consider 1-dimensional layers, 
and use the intra-layer hopping $t^\pp=1$ as the unit of energy. The equilibrium spectral function for an infinite system of 
such 1-$d$ layers and inter-layer hopping $\tp=1$ (corresponding to the usual 2-$d$ Hubbard model) is shown in Fig.~\ref{equilibrium}, 
for inverse temperature $\beta=5$ and indicated values of $U$. The impurity problem was solved with NCA on the Keldysh contour and the 
spectra were obtained via Fourier transformation of the retarded Green's function. Around $U=7$, a Mott gap opens in a continuous 
fashion (crossover). Since we cannot reliably study the 
low temperature behavior of the metallic phase
within NCA, we will not investigate this transition  
in further detail.
In the following, we will mostly focus on the insulating regime ($U>7$).

A good test of the implementation and its accuracy is the calculation of the total energy. The total energy, normalized by the number of 
sites in the transverse direction, has a local contribution 
\begin{equation}
E_\text{pot}=\sum_{m=1}^{N}[U_m d_m + (\epsilon_{\text{loc},m}-\mu )n_m],
\end{equation} 
where $d_m$ is the double occupancy and $n_m=n_{m\uparrow}+n_{m\downarrow}$ the occupation on layer $m$.  In addition there is 
the intra-layer kinetic energy
\begin{equation}
E_\text{kin,intra}=\sum_{m=1}^N 
\sum_{\kk\sigma}
\epsilon_{\kk,m} n_{\kk,m,\sigma},
\end{equation} 
and the inter-layer kinetic energy  
\begin{equation}
E_\text{kin,inter}=-\sum_{m=1}^{N-1} 
\sum_{\kk\sigma}
\,t^\zz_{m}\, \langle c^\dagger_{\kk,m,\sigma}c_{\kk,m+1,\sigma}\rangle + h. c.,\label{ekin_inter}
\end{equation} 
where we have assumed vacuum boundary conditions. 
To evaluate Eq.~(\ref{ekin_inter}) we note that 
$t^\zz_{m} \langle c^\dagger_{\kk,m,\sigma}c_{\kk,m+1,\sigma}\rangle
=-it^\zz_{m}G_{\kk,m+1,m}^<(t,t)$ and $t^\zz_{m} G_{\kk,m+1,m} = \Delta^R_{\kk,m+1} \ast G_{\kk,m}$. The latter identity follows 
from a comparison of the Dyson equation (\ref{dyson-chain}) with Eq.~(\ref{Gknn}).

\begin{figure}[t]
\begin{center}
\includegraphics[angle=0, width=0.88\columnwidth]{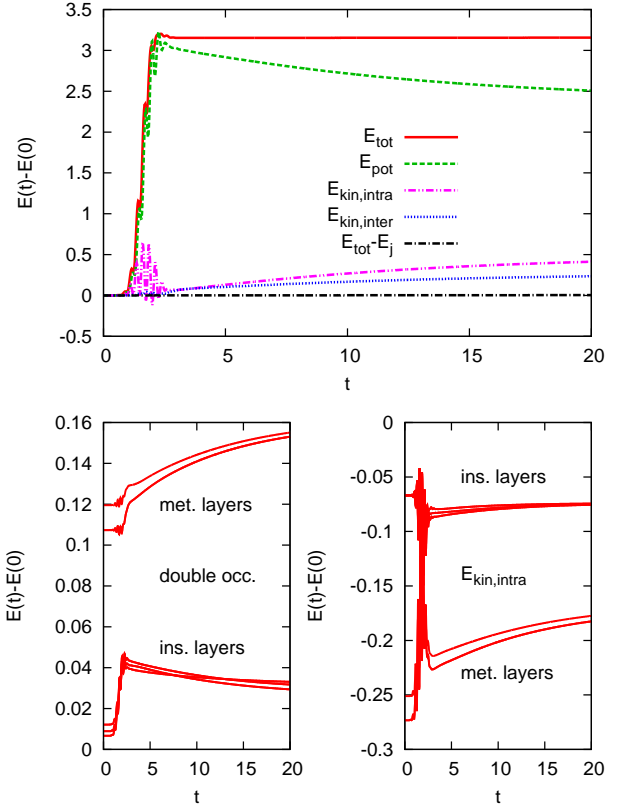}
\caption{Test of the energy calculation for a heterostructure composed of nine 1-$d$ layers with $t^\perp=1$, interaction $U=15$ on layers 
1, 2, 3, 7, 8, 9 and $U=4$ on layers 4, 5, 6, with ``vacuum" boundary conditions. An electric field pulse is applied to all the layers.
}
\label{energy_test}
\end{center}
\end{figure}

\begin{figure}[t]
\begin{center}
\includegraphics[angle=0, width=\columnwidth]{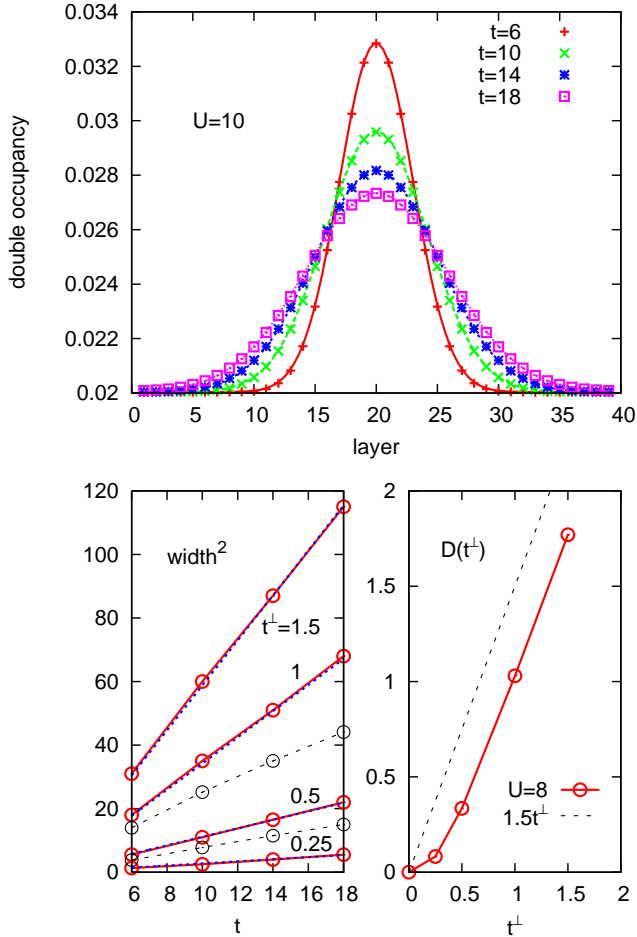}
\caption{
Top panel: Time-dependent distribution of the double occupancy in a 39-layer system with $U=10$, after a pulse excitation with $\Omega\approx 12$ applied to the middle layer. Lines show a fit to a Gaussian centered at the middle layer. The bottom left panel plots the widths extracted from such fits as a function of time for different values of the inter-layer hopping. The squared width grows linearly in $t-t_\text{pulse}$, where $t_\text{pulse}\approx 1.7$ is the time corresponding to the center of the pulse. Dashed lines show results with phonon bath (see text). The bottom right panel shows the dependence of the diffusion constant $D$ on the inter-layer hopping. 
}
\label{gauss}
\end{center}
\end{figure}

If an electric field is applied to the system, a current $j$ 
will be induced ($j$ is defined as 
the particle current, not including the electric charge $-1$),
\begin{align}
j^\pp_m
&= 
\sum_{\kk\sigma}
(\partial_\kk \epsilon_{\kk,m}) n_{\kk,m,\sigma},
\\
j^\zz_m 
&= 
-i\,
\sum_{\kk\sigma}
t^\zz_{m} \langle c^\dagger_{k,m,\sigma}c_{k,m+1,\sigma} \rangle - h.c.,
\end{align}
where  $j^\pp_m$ is the intra-layer component, and $j^\zz_m$ is the current from layer $m$ to $m+1$. 
% PW: would it be more natural to write "current from layer $m+1$ to $m$"? 
%ME I should have continuity equation like d/dt n_m = j_{m-1}-j_m if j_m is "the current from m to m+1" 
While the electric field is applied, the total energy will change like 
$d E_\text{tot} /dt = -\sum_m{\bm j}_m{\bm E}_m$  
(for electrons with charge $-1$).
Here we assume vacuum boundary conditions (and thus $j^\zz_0=j^\zz_N=0$), because otherwise 
energy can flow from the system into the leads.
A good check of the numerics 
is thus to verify that $E_\text{tot}(t)-E_j(t)$ is time independent, 
where $E_j(t)=
-\int_0^t\!d\bar t
\big[\sum_{m=1}^N
j^\pp_m(\bar t)
E^\pp_m(\bar t) +
\sum_{m=1}^{N-1} j^\zz_m(\bar t) E^\zz_m(\bar t)\big]$ 
is the absorbed energy.
After the pulse, the Hamiltonian of the system is time-independent, and the total energy should 
thus also become time-independent. 
In Fig.~\ref{energy_test} we plot the time-evolution of the different 
energy contributions
for a nine-layer system consisting of three metallic layers ($U=4$) sandwiched between Mott insulating layers ($U=15$). 
The perturbation is an in-plane, few-cycle electric field pulse of frequency $\Omega\approx 12$, which is applied to all nine 
layers.  This strong pulse creates doublon-hole pairs and leads to a rapid increase in the potential energy. After the pulse, 
one observes a redistribution of potential energy into kinetic energy in such a way that the total energy is conserved. Also, 
the change in total energy is equal 
to the absorbed energy 
$E_j$, so that 
$E_\text{tot}-E_j$ 
remains zero within the numerical 
accuracy. That this result is a nontrivial check follows from the lower panels, which show the time-evolution of the double 
occupancy and intra-layer kinetic energy in all nine layers. These curves indicate that doublons and holes move from the 
insulating regions to the metallic region, where they recombine, heat up the metal and lead to an increase in the intra-layer 
kinetic energy.

\subsection{Doublon diffusion}

As a first application we consider the  spreading of photo-excited doublons and holes in a Mott insulator. The system 
consists of 39 layers and we employ the ``repeated" boundary condition to minimize boundary effects. The doublons 
and holes are created in the central layer ($m=20$)  by the application of an in-plane electric field pulse with 
$\Omega\approx 12$, centered at $t_\text{pulse}=1.7$, which lasts up to $t=3$. This set-up may not be realistic 
from an experimental point of view, but it allows us to study how artificially created carriers spread out inside a 
Mott insulating bulk. On the timescale of the present simulation, we can ignore the recombination of doublons and holes. This 
is consistent with corresponding DMFT calculations for a homogeneously excited bulk system, which indicate that 
the lifetime of these carriers depends exponentially on the interaction $U$ in the Mott insulating regime.\cite{Eckstein2011} 
As shown in the top panel of Fig.~\ref{gauss} (results for $U=10$), already a short time after the pulse, the 
distribution of the photo-excited doublons (symbols) can be well fitted by a Gaussian (lines). 
For inter-layer hopping $\tp=1$, the 39-layer system allows us to track the motion of the doublons up to 
$t\approx 20$. 
Extracting the widths of the Gaussians and plotting them as a function of time (Fig.~\ref{gauss}, lower left panel), 
we find that the square of the width grows proportional to $t-t_\text{pulse}$, indicating diffusive rather than ballistic motion.
The doublon diffusion satisfies the expected law  
$d(m,t)-d_\text{eq}(m,t)\sim \exp(-(m-20)^2/(4Dt))$
($m=20$ is the 
central layer), with diffusion constant $D\approx 1.03 \tp$ for $\tp=1$. 
As long as the doublon-holon recombination is slow enough and the carriers are inserted with large kinetic 
energy, the diffusion of doublons and holes is not influenced much by the interaction strength. 
Within our numerical accuracy, we find the same diffusion constant for $U=7$, $8$, $9$ and $10$, even though 
$U=7$ is already close to the metal-insulator crossover. 

On the other hand, a smaller inter-layer hopping of course slows down the diffusion. The lower left panel of 
Fig.~\ref{gauss} plots the time-evolution of the squared width of the distribution, for $t^\zz$ ranging from 
$0.25$ to $1.5$ ($U=10$).
(Because of the rapid spreading of the charge carriers we cannot 
study much larger values of $t^\zz$.)
The diffusion constant $D(t^\zz)$, which is extracted from linear fits to these curves, grows roughly quadratically 
with $t^\zz$ for small $t^\zz$, while the dependence becomes almost linear for $t^\zz\gtrsim 0.5$
(Fig.~\ref{gauss}, lower right panel). 

\begin{figure}[ht]
\begin{center}
\includegraphics[angle=0, width=\columnwidth]{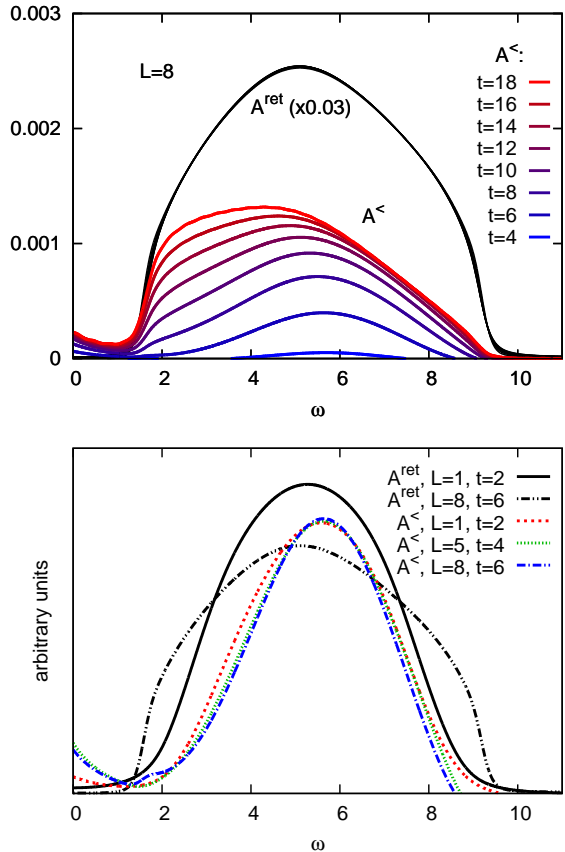}
\caption{
Spreading of doublons in a 15-layer system with $U=10$ and pulse-excitation with $\Omega\approx 12$ on layer 1.  
The top panel shows the increase of the occupied part of the spectrum in layer 8, and the relaxation of the carriers towards the bottom of the upper band.
The bottom panel shows the spectral functions on layers 1 and 8 (solid and dashed black lines) and the occupied part of the spectrum in layers 1, 5 and 8 a short time after the injection of carriers (rescaled in such a way that the maxima are approximately the same).  The distribution of the fastest carriers remains almost unchanged from layer 1 to 8.   
}
\label{doublon spreading 1}
\end{center}
\end{figure}

In equilibrium, the diffusion constant is related to the conductivity $\sigma_\text{dc}$
(with the charge $e$ set to one) and the compressibility $\frac{\partial n }{\partial \mu}$ via the Einstein relation 
(or fluctuation-dissipation relation)
\begin{equation}
\label{einstein}
D \frac{\partial n }{\partial \mu} = \sigma_\text{dc} .
\end{equation}
Truly ballistic transport is thus expected for integrable one-dimensional systems (see Ref.~\onlinecite{Sirker2009} 
and references therein), which can have a perfect conductivity (i.e., a finite Drude weight $\sigma_\text{dc} \sim 
\mathcal{D}\delta(\omega)$ for $\omega\to0$) even at temperature $T>0$.\cite{Zotos1997}  For the Hubbard model 
in higher dimensions, the rather large width of the spectral function $A_{\kk\sigma}(\omega)$ in the Mott insulator 
indicates that the scattering time of a single particle excitation with momentum $\kk$ is of the order of the inverse 
hopping, and hence its mean free path is not much larger than a few lattice spacings. Only in a Fermi liquid at 
$T=0$ would one expect infinite scattering times for electrons at the Fermi surface.

To some extent, the behavior of $D(t^\zz)$ shown in the lower right panel of Fig.~\ref{gauss} is qualitatively 
consistent with a quasi-equilibrium argument based on the Einstein relation for large temperature $T$: 
Starting from the DMFT expression for the bulk conductivity,\cite{Georges1996,Pruschke93} 
\begin{multline}
\label{sigma-eq}
\sigma_{\alpha\alpha'}
(\omega) 
\propto
\sum_{\kk\sigma}
v_{\kk}^{\alpha}
v_{\kk}^{\alpha'}
\int_{-\infty}^{\infty} \!d\omega'\,
\times
\\
\times
\frac{A_{\kk\sigma}(\omega')A_{\kk\sigma}(\omega+\omega')
[f(\omega')-f(\omega+\omega')]}{\omega},
\end{multline}
($\alpha=\zz,\pp$), 
the dc conductivity in the transverse direction and in the limit of high temperature 
is given by
\begin{equation}
\label{sigma-eq}
\sigma^\zz_\text{dc}
\propto
\frac{1}{4T}
\sum_{\kk\sigma}
(v_{\kk}^\zz)^2
\int_{-\infty}^{\infty} \!\!\!d\omega\,
A_{\kk\sigma}(\omega)^2,
\end{equation}
where $v_{\kk}^\zz=t^\zz \sin(k^\zz)$ is the band velocity perpendicular to the layers. The integral scales 
like $1/$bandwidth, 
and the bandwidth is proportional to $t^\pp$ for $t^\pp \gg t^\zz$ (almost independent layers), 
and proportional to $t^\zz$ for $t^\zz \gg t^\pp$. Thus $\sigma_\text{dc}^\zz\sim t^\zz$ for  
$t^\zz \gg t^\pp$ and $\sigma_\text{dc}^\zz\sim |t^\zz|^2/t^\pp$ for  $t^\pp \gg t^\zz$. 
Because $\frac{\partial n }{\partial \mu} \sim n/4T$ for large $T$, the 
same behavior is found for the diffusion constant $D(t^\zz)$. Physically, the behavior for 
small $t^\zz$ is consistent with a rate equation picture, where the transfer of a doublon from 
one layer to the next is given by Fermi's golden rule $\Gamma \sim |t^\zz|^2 \mathcal{N}$, 
with a matrix element $\propto t^\zz$, and a density of states $\mathcal{N} \sim 1/t^\pp$ that 
scales with the inverse bandwidth. 

Although the Einstein relation agrees with the observed behavior on a qualitative level, 
such a  quasi-equilibrium theory cannot describe the spreading of doublons in detail. 
First of all, the initial perturbation of the system is strong, and it is neither clear on what timescale 
a local equilibrium description becomes possible, nor how well it would apply to a distribution that 
varies considerably over only a few lattice spacings. 
Since doublons and holes might cool down (lower their kinetic energy) while they
spread in the bulk, equilibration could actually lead to the formation of Fermi liquid quasi-particles
and a corresponding reconstruction of the electronic density of states, a process for which the 
time-scale is not known.  Examples where nonequilibrium conditions have a strong influence on 
the spreading of particles have been studied recently, for a cloud of weakly-interacting ultra-cold 
atoms in an optical lattice (both fermions  and bosons).\cite{Mandt2011,Schneider2012,Ronzheimer2013} 
For example, when the cloud expands into an empty lattice, it behaves diffusive in the dense core, 
but in the tails the density is too low to equilibrate, resulting in a ballistic expansion.\cite{Schneider2012,Ronzheimer2013} 

More detailed insight into the way in which doublons and holes spread into the bulk can 
be obtained from the time- and layer-dependent distribution function
\begin{equation}
A^<_m(\omega,t)
=
\frac{1}{\pi}\text{Im}
 \int_0^\infty
\!ds\, 
e^{i\omega s}\,G_m^<(t+s,t),
\label{distribution N}
\end{equation}
which reduces to the ``photoemission spectrum'' $A^<(\omega,t)=A(\omega)f(\omega)$ 
in equilibrium, and from the corresponding spectral function $A_m(\omega,t)$ (with $G^<$ 
replaced by $-G^R$). 
To study this quantity we switch to a smaller system, so that 
longer simulation times become possible and
the integral in Eq.~(\ref{distribution N}) 
does not strongly depend on to the upper cutoff. In the upper panel of Fig.~\ref{doublon spreading 1}, 
we plot the distribution function for a 15-layer system with $U=10$, which is excited with a pulse with 
$\Omega\approx 12$ on the surface layer $n=1$. (A ``repeated'' boundary condition is applied at layer 15.) 
On a given layer $L$, ($L=8$ is plotted in the figure), the weight in the upper Hubbard band grows with 
time as more doublons arrive. At later times, the distribution is shifted to lower frequencies, indicating 
some kind of cooling of the particles as they move into the bulk. Still, the distribution is clearly 
non-thermal at all times, and its width remains comparable to the width of the Hubbard band. In such a highly
excited system, one cannot expect the formation of quasi-particle states. Indeed, we only observe a slight 
broadening of the spectral function, rather than a formation of a quasi-particle band. 

Although the weight in the distribution function $A_m^<(\omega,t)$ appears after an increasing-time delay 
as one moves further away from the surface, we find that the the distribution at the earliest times
(i.e., right after it has achieved some measurable weight) has a similar shape on different layers 
(Fig.~\ref{doublon spreading 1}, lower panel). The distribution resembles  the initial photo-doped 
distribution on layer $1$, although the spectral function of the bulk layers is quite different from 
that of the surface layer, 
especially during the application of the pulse. This might be related to a coherent tunneling at early times. 

A detailed understanding of the various propagation effects at early and later times can be important to 
interpret the relaxation of photo-excited carrier distributions in real experiments, which is governed 
by both diffusion and local relaxation phenomena. In real materials, doublons and holes can
dissipate their energy to other degrees of freedom as they diffuse into the bulk, e.g., to phonons 
or spin excitations, which are not correctly accounted 
for in the DMFT formalism
for the isolated Hubbard  model.
 To study the consequences of 
 this dissipation,
 we have simulated the diffusion in 
the presence of a local phonon bath with $\omega_0=1$ and $\lambda=1$. In this case the doublons 
and holes spread more slowly, as shown for $\tp=1$ and $\tp=0.5$ by the dashed lines in Fig.~\ref{gauss}. 
A possible explanation is that the phonon cloud increases the effective mass of the carriers 
and hence reduces their diffusion coefficient. On the other hand, the curve for 
$\tp=1.0$ also reveals a slight negative curvature, which indicates that the cooling of the carriers 
influences
the diffusion behavior in a nonlinear way.

\subsection{Surface excitation of a heterostructure and doping by diffusion}

\begin{figure}[t]
\begin{center}
\includegraphics[angle=0, width=\columnwidth]{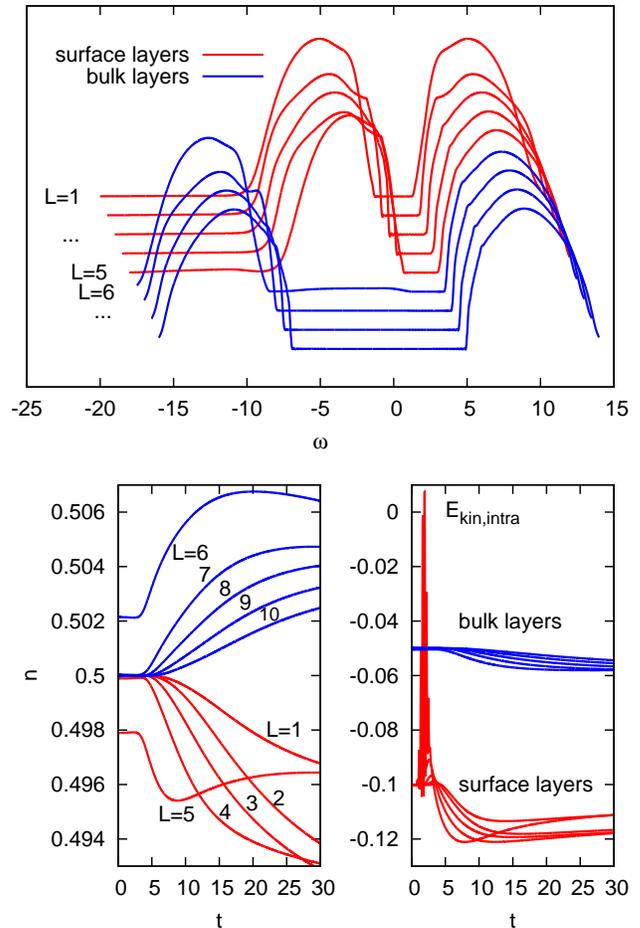}
\caption{
Top panel: Spectral functions for a heterostructure with five small-gap insulating layers (red, $U=10$) on top of a bulk of large-gap insulator (blue, $U=20$). The bands are shifted relative to each other in such a way that doublons can diffuse from the surface into the bulk, while holons cannot. The spectra are plotted with horizontal offsets of $0.3$ and arbitrary vertical offsets. Only four out of ten bulk layers are shown. 
Bottom panels: Time evolution of the filling and intra-layer kinetic energy after excitation of the surface layer with a $\Omega\approx 12$ field-pulse on the surface layer. (Phonon coupling $\lambda=1$, phonon-frequency $\omega_0=1$, $\beta=10$.)
}
\label{heterostructure}
\end{center}
\end{figure}

\begin{figure}[t]
\begin{center}
\includegraphics[width=\columnwidth]{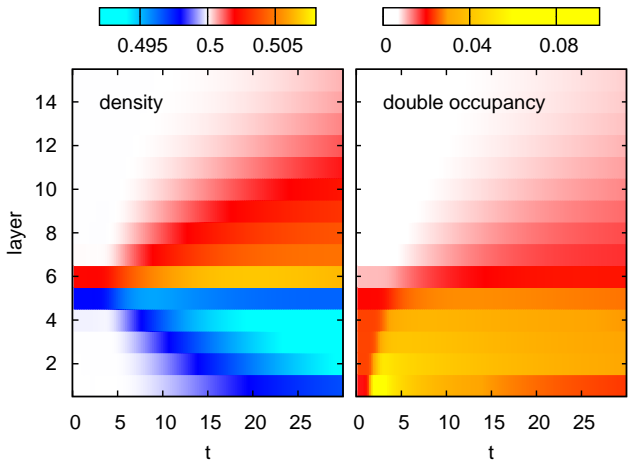}
\caption{
The left panel shows the time-evolution of the density, and the right panel the time-evolution of the double occupancy in the set-up of Fig.~\ref{heterostructure}.
}
\label{heterostructure_n}
\end{center}
\end{figure}

\begin{figure*}[ht]
\centerline{\includegraphics[width=0.8\textwidth]{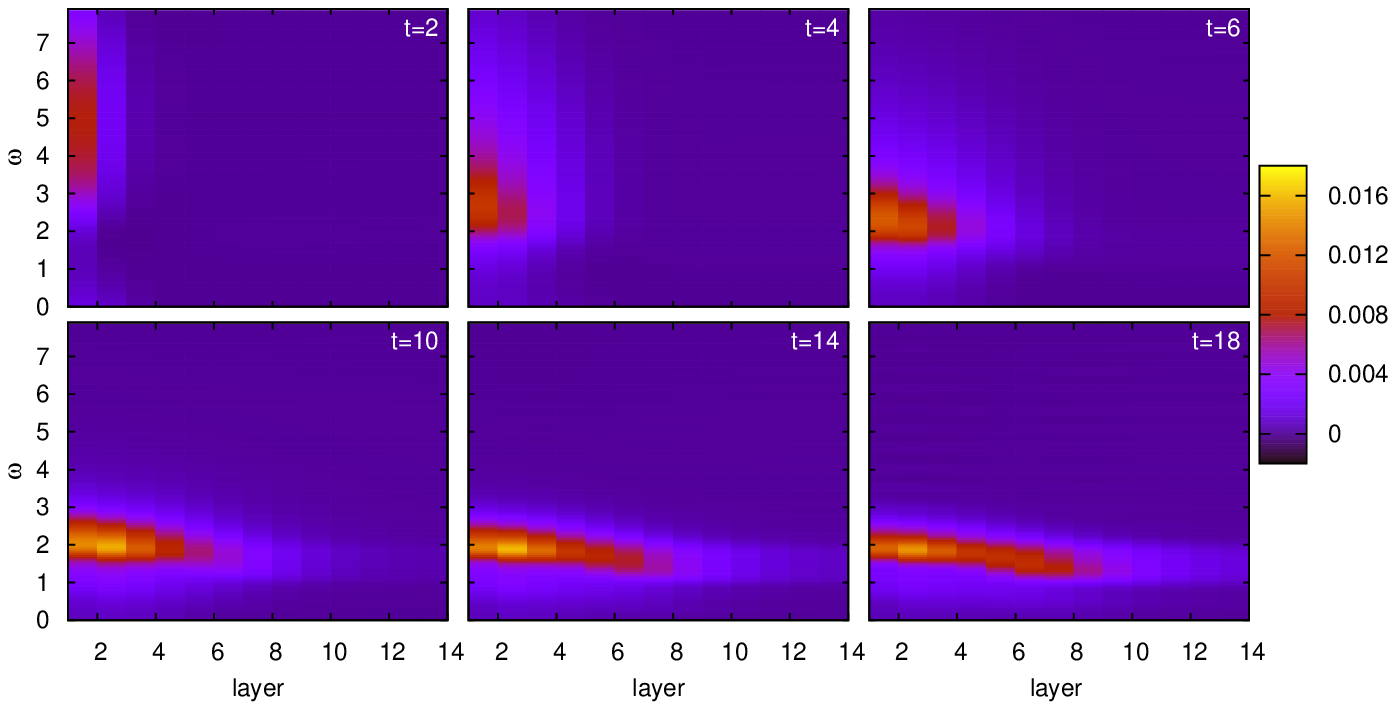}}
%\centerline{\includegraphics[width=0.8\textwidth]{doublon_3d_new.eps}}
\caption{
Occupation function $A^<(\omega,t)$ in the upper Hubbard band, for the same set-up as Fig.~\ref{heterostructure}.
The excitation is an in-plane field pulse with frequency $\Omega \approx 12$ on the surface layer. 
}
\label{diffusion}
\end{figure*}

An interesting 
application of the layer DMFT 
is to study the dynamics in heterostructures. 
Experimentally, such artificially designed systems may provide a way to confine the 
excitation to a well-defined region of the sample (because, e.g., the pulse frequency can be tuned to the absorption band 
in certain layers), and induce controlled changes in the remaining layers. For illustration, we consider a 
heterostructure made of two different Mott insulators, and excite doublons and holes in the topmost layer. 
As illustrated in the top panel of Fig.~\ref{heterostructure}, the system consists of five Mott insulating surface layers (red spectral functions) on top of a Mott insulating bulk, whose gap is much larger than the gap of the surface layers (blue spectral functions). The relative position of the Hubbard bands is chosen such that doublons 
can diffuse easily from the surface layers into the bulk, while the corresponding diffusion of holes into the bulk is prohibited. 

The diffusion of charge carriers leads to a time-dependent doping of the neighboring layers with electrons and holes,
and the special setup of Fig.~\ref{heterostructure} allows to study the possible time-resolved emergence of a usual metallic state in the bulk
layers, which are doped with electrons only. Explicitly, we simulated five surface layers with $U=10$ on top of 
ten bulk layers with $U=20$. We choose the ``vacuum" boundary condition for the surface layer $n=1$, and apply 
the in-plane electric field to this layer.  To mimic dissipation to lattice and other degrees of freedom, which can 
accelerate the formation of a photo-doped state with low kinetic energy and less scattering, we couple the 
system to local phonon baths, as described in the methods section and in Ref.~\onlinecite{Eckstein2012photodoping}. 
The phonon bath parameters are $\omega_0=1$ and $\lambda=1$. 
(The small structures visible in the spectral functions near the gap edges are a result of this phonon coupling.)

The electron doping of the bulk- and net hole doping of the surface layers can be seen in the bottom left panel of Fig.~\ref{heterostructure}, 
which plots the time-evolution of the density for the different layers.  Note that even in the equilibrium system, a charge transfer occurs at 
the interface between surface and bulk layers, so that the first bulk layer is $0.2\%$ electron doped, while the last surface layer is $0.2\%$ 
hole doped. Figure~\ref{heterostructure_n} shows the time-evolution of the double occupancy and the density on a color scale 
(grey scale).
Initially the double occupancy is slightly larger in the surface layers, due to the smaller value of $U$.
We find that the interface between the two insulating regions does not slow down the diffusion of doublons into 
the bulk layers, while holes stay confined to the surface layers. 
There is even an accumulation of doublons on the bulk side of this interface, 
which is explained by a small downward shift of the Hubbard band. 
The net charge in the surface layers is reduced as time increases  due to the holes which diffuse back from the interface. 

As a result of the dissipation, we expect the doublons and holes, which are created in the Hubbard bands of the $L=1$ layer with 
a broad energy distribution, to cool down rapidly while they diffuse into the bulk. The latter effect should be evident as an 
accumulation of spectral weight in the distribution function (\ref{distribution N}) at the lower edge of the upper Hubbard band 
(and symmetrically for the holes). 
Figure~\ref{diffusion} illustrates the time-evolution of the occupied spectral function in the upper Hubbard band, which roughly covers the energy range 
$1.5\lesssim \omega\lesssim 10$. The few-cycle pulse with $\Omega\approx 12$ creates doublons with a broad energy distribution centered at 
$\omega\approx 6$ (in the middle of the upper band). Such a broad spectrum is visible in the surface layer at $t=2$ (the pulse lasts from about $t=0.4$ to 
$t=3$). Very quickly (top right panel, $t=4$), the doublons spread to the neighboring layers, and the cooling by the phonon bath leads to a shift of spectral 
weight to lower energies. Around $t=6$, the diffusing doublons reach the bulk layers ($n\ge 6$). They keep diffusing into the bulk, which results in a pure 
electron doping of the bulk layers. Furthermore, by $t=10$, the phonon bath has removed most of the excess kinetic energy so that the changes in the 
spectral function at later times are mainly due to changes in the carrier density.

The decrease in the total in-plane kinetic energy in the different layers is also evident in the bottom right panel of Fig~\ref{heterostructure}.
This is consistent with a a metallization of the bulk layers as a result of the doping induced by the diffusion of doublons. The 
small quantitative change of the kinetic energy is explained by the small amount of doping and the high effective temperature 
of the doped system. Despite the strong coupling to the phonon bath with inverse temperature $\beta=10$, the distribution 
function $A^<(\omega,t)$ remains non-thermal within the accessible time-range, and it is much broader than expected for 
the effective temperature of the bath. In addition, no pronounced quasiparticle peak emerges in the spectral
function on these timescales. 
As in the case of a photo-doped metallic state with electrons and holes,\cite{Eckstein2012photodoping} it seems that the 
purely electron doped state obtained via doublon diffusion from the surface layer is not a good metal, and that the formation 
of a Fermi liquid state similar to an equilibrium chemically doped Mott insulator is a very slow process. 

Finally we
note
that in principle one should consider also the electrostatic energy associated with the (time-dependent) charge redistribution. This could be done for example by adding a layer-dependent Hartree potential $V_m(t)$ to the chemical potential in the DMFT loop. The formulas for this potential are given, for example, in Refs.~\onlinecite{Oka2005, Charlebois2012}:
\begin{equation}
V_m(\{n_1, \ldots, n_{m-1}\},t) = -\alpha\sum_{k=1}^{m-1}\sum_{l=1}^k (n_l(t)-n_\text{background}),
\end{equation}
where $n_\text{background}=0.5$ for half-filling and $\alpha$ is a constant proportional to the inverse dielectric constant.  
This potential would stop the spreading of charge into the bulk and confine the carriers to a region close to the interface. However, since the main purpose of the present work is to explain the nonequilibrium real-space DMFT method and to illustrate its versatility with several examples, we will leave the calculation of realistic time-dependent charge profiles in heterostructures to a future publication. (The results shown here are representative of materials with a large dielectric constant.)

\begin{figure*}[ht]
\begin{center}
\includegraphics[angle=0, width=1.1\textwidth]{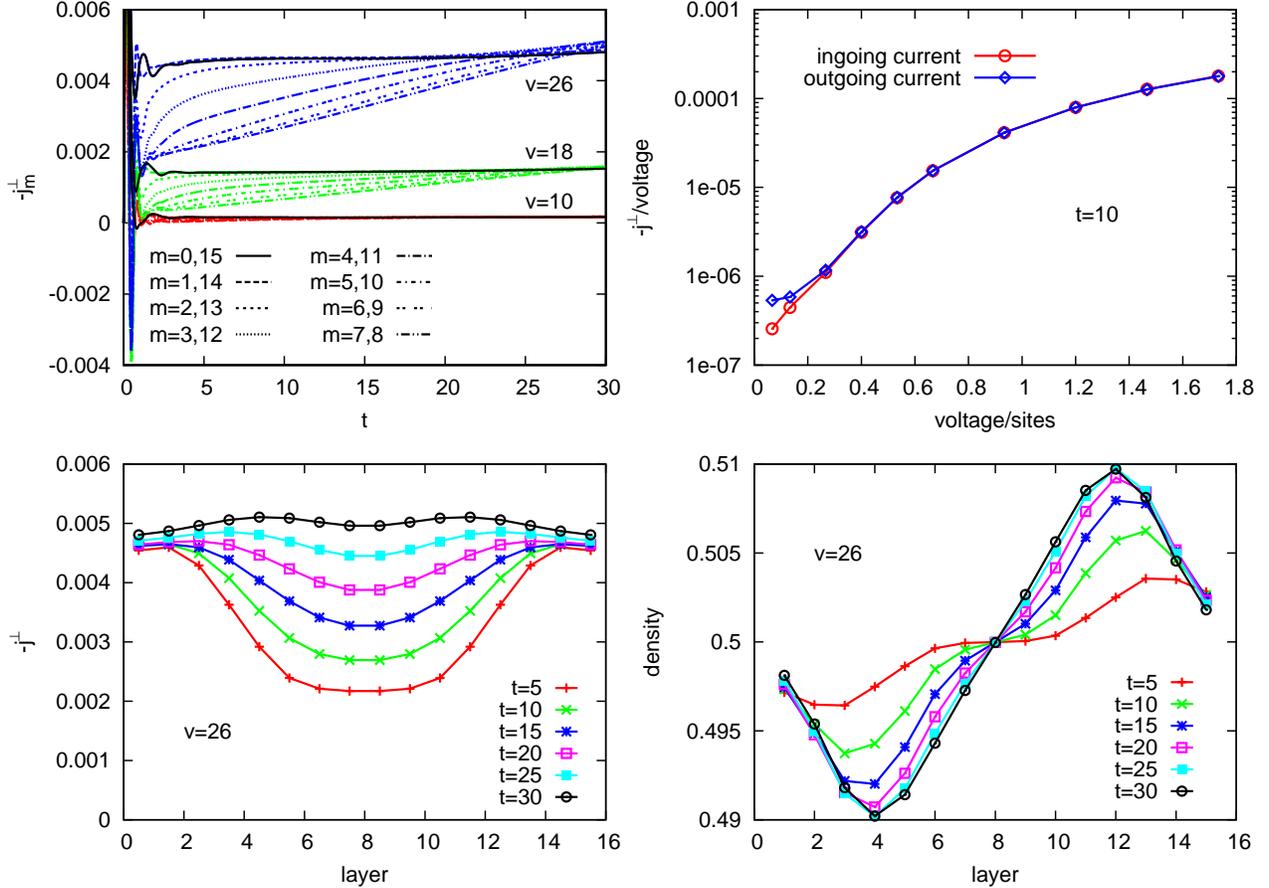}
\caption{
Top left panel: Current between the different layers of a 15-layer system with $U=10$ as a function of time after the switch-on of a voltage bias $v$. The black line shows the current which flows into the system, which at the resolution of the plot is indistinguishable from the current which flows out. 
Top right panel: Current measured at the leads at $t=10$, divided by voltage and plotted as a function of voltage, showing an exponential increase at low bias. The slope gives the threshold voltage for the dielectric breakdown of the Mott insulator.
Bottom panels: Current and charge distribution for different times. 
}
\label{current}
\end{center}
\end{figure*}

\begin{figure}[ht]
\begin{center}
\includegraphics[angle=0, width=\columnwidth]{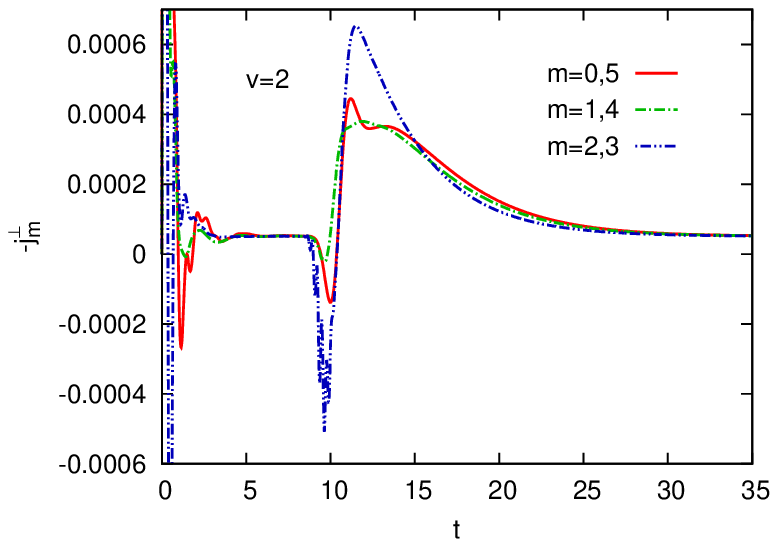}
\caption{
Current induced in a Mott insulating 5-layer system ($U=10$) under a voltage bias $v=2$, by an in-plane field pulse acting on the middle layer. The pulse with $\Omega\approx 12$ is centered at $t_\text{pulse}=9.3$. Initially, the system is in equilibrium, and the large current spikes at short times are due to the build-up of a polarization after the switch-on of the linear bias. 
}
\label{current_pulse}
\end{center}
\end{figure}

\subsection{Multi-layer structures under applied bias}

Transport through nanoscopic devices is another important area of physics that involves both nonequilibrium 
phenomena and strong correlations. The nonlinear current voltage characteristics of a two-terminal 
heterostructures has been studied previously, using an inhomogeneous steady-state DMFT approach.\cite{Okamoto2008a} 
The present formalism allows to study such systems in real time, and as a first application, we investigate the 
time-dependent build-up of current and charge distributions across the sample after the switch-on of a voltage-bias perpendicular to the layers.
We consider a system consisting of $L=15$ correlated layers in the Mott regime ($U=10$). In these calculations we do not attach 
local heat baths, so that energy dissipation occurs only in the leads, and may not be relevant on the time-scales of our simulations. Initially, the 
system is in equilibrium without applied bias, and at time $t=0$, we switch on a bias $v$ across the whole sample, assuming that the voltage 
drop is linear, i.e., the electric field is $E^\zz =v/(N+1)$. 
The top left panel of Fig.~\ref{current} shows the time-evolution of the current 
$j^\zz$ flowing between the layers, for three different values of $v$.
After some initial strong oscillations of the current $j^\zz$, which are related to the build-up of a polarization perpendicular 
to the layers, the currents into layer $1$ 
($j_0^\zz$) 
and out of layer $15$  
($j_{15}^\zz$)
quickly settle to some $v$-dependent value which changes only slowly with time (bold lines). This is in contrast to the currents between layers in the interior 
of the sample, which show a slower time evolution and no relaxation into a quasi-steady state up to $t=30$. The almost steady currents into and 
out of the leads exhibit a similar threshold behavior as was found in single-site DMFT calculations,\cite{Eckstein2010breakdown, Eckstein2013breakdown} 
i.e., an exponential increase at low bias of the form $j^\zz\propto v \exp(-v_\text{th}/v)$. This is illustrated in the top right panel of 
Fig.~\ref{current}, which 
plots the ingoing and outgoing current at time $t=10$ on a logarithmic scale. 

In the bottom left panel
of Fig.~\ref{current}
 we show current profiles  within the structure at different times, for $v=26$. At short times, the current is largest near the 
leads and smallest in the center. Around $t=27$, the current deficit in the center changes into a current surplus (see also upper left panel), and 
we can expect some oscillations, until eventually an almost flat quasi-steady state distribution is established. This current profile implies a 
redistribution of charge from the left side of the multi-layer structure to the right side at short times. Indeed, a similar plot of the density 
distribution (bottom right panel) shows a build-up of positive (negative) charge in the left (right) half of the structure which progresses 
from the boundaries. At $t=30$ the excess charge peaks at layers $4$ and $12$, which is in the middle of the left and right regions. 
The distribution in the quasi-steady state might look similar. Again, one should in principle take the electrostatic potential associated 
with this charge redistribution into account and compute the potential profile across the structure self-consistently. 

Similar time- and layer-dependent redistribution processes might be observable if they are triggered by a short pulse.
To illustrate this, we finally discuss
the current induced in Mott insulating structures under bias by an applied intra-layer electric field pulse. We consider a 
$5$-layer structure with $U=10$. The voltage $v=2$ across the insulating sample is small enough that after the build-up 
of a polarization, there is 
only a very small 
current flowing through the sample 
(Fig.~\ref{current_pulse}). 
Between 
$t=8$ and $t=11.6$ a field pulse with $\Omega\approx 12$ is applied to the middle layer (with polarization in the in-plane direction).
At later times, the doublons and holes created by the pulse start to diffuse to the leads under the applied bias, 
which leads to a net negative current. 
The decay of this current  is a direct measure for the mobilities.
The intra-layer current during the pulse 
exhibits a peak in the opposite direction to the expected bias-induced current 
in the central region, which indicates that the polarization in the central layers is reduced in response to the 
perturbation.

\section{Conclusions}
\label{Sec_conclusions}

We have described and tested the nonequilibrium extension of real-space DMFT, which allows to study layered systems with strong 
electronic correlations. Like single-site DMFT (and in contrast to cluster-extensions of DMFT), the formalism is based on the assumption 
of a purely local self-energy. One thus only has to solve a collection of (coupled) single-site impurity problems in a self-consistent manner. 
For a layer geometry, in which all properties of the system depend on only one space direction, the computational effort scales linearly 
with system size (up to the number of iterations, which may weakly depend on the system size), and the same is true for the storage 
requirement. We have discussed the details of our implementation based on self-consistent strong-coupling perturbation theory (NCA) 
as an impurity solver, but the formalism can equally be combined with a Monte Carlo,\cite{Werner2009} or a perturbative weak-coupling solver.\cite{Eckstein2010}

As an application, we have simulated the diffusion of photo-excited doublons in a Mott insulator, both inside the bulk, and 
from the surface of a heterostructure into the bulk. 
The diffusion constant was found to depend mainly on the inter- and and intra-layer hopping, while 
it is almost independent of the interaction strength. 
A heterostructure set-up allows for a controlled doping of charge carriers of one type (e.g., doublons) into a Mott insulator, 
in contrast to photo-doping, where always both electrons and holes are inserted. In principle, this opens
the possibility to study the formation of quasi-particles in a metallic system. For the current set-up, however, 
we find that the timescale for the build-up of such a state is rather long, such that the doped system behaves
more like a bad metal on the numerically accessible timescales. A more thorough investigation of this 
important question will be deferred to a future study.

The second type of application was the layer- and time-resolved calculation of the current through a correlated 
insulating slab, where we reproduced the threshold behavior of the current-voltage characteristics known from 
previous nonequilibrium DMFT studies, and computed the evolution of the current- and density-profile after the 
switch-on of the voltage bias. We also studied a Mott insulating slab under bias (below the threshold for the dielectric 
breakdown) where the time-dependent redistribution of charge after a few-cycle laser pulse can be studied.

In the future, one should include the effect of the electrostatic potential to obtain a more realistic description of the 
diffusion of electrons and holes in a heterostructure. 
Also, the extension of our formalism to antiferromagnetically ordered layers would be useful, 
because this would allow to exploit the cooling effect 
on the photo-doped carriers
associated with demagnetization.\cite{Werner2012} 
How, and on which time-scale an almost thermal metallic state can be induced in a Mott insulator by diffusion of 
doublons from neighboring layers is an interesting topic for further studies.

\acknowledgements 
We thank H. Aoki, T. Oka and N. Tsuji for helpful discussion. 
PW acknowledges support from FP7/ERC starting grant No. 278023.

\end{document}